\documentclass{aastex62}

\shorttitle{Formation of SS433}
\shortauthors{}

\begin{document}

\title{On the Formation of SS433}

\author{Qin Han and Xiang-Dong Li}
\affil{Department of Astronomy, Nanjing University, Nanjing 210023, China;
lixd@nju.edu.cn}

\affil{Key Laboratory of Modern Astronomy and Astrophysics, Nanjing University,
Ministry of Education, Nanjing 210023, China}

\begin{abstract}
SS433 is an extraordinary X-ray binary which is ejecting bipolar jets with $26\%$ of the speed of light. Associated with the supernova-like shell W50, SS433 is also probably one of the youngest X-ray binaries with an age of $\lesssim 10^5$ yr. However, the masses of the two components in SS433 and even the nature of the compact object are still under debate. In this work, assuming that the compact object is a black hole, we employ a binary population synthesis method to study the formation of SS433. We use previous estimates of the age of W50 and the duration of the jet activity to constrain the evolutionary history. Our calculations suggest that SS433 likely harboured a Hertzsprung gap star at the beginning of the current Roche-lobe overflow phase. The masses of the black hole and the optical/donor star in SS433 suggested by the simulations are around $8\ M_\sun$ and $24\ M_\sun$, respectively. Future measurement of the donor mass and radius can help infer not only the orgin of the binary but also the nature of the nebula W50.
\end{abstract}


\section{Introduction} \label{sec:intro}

SS433 is a Galactic X-ray binary (XRB) that has been intensively studied. It has some unique properties. Perhaps the most remarkable one is the two precessing, mildly relativistic ($0.26$ speed of light) jets \citep{1979MNRAS.187P..13F,1979A&A....76L...3M,1979Natur.279..701A}. Mass is also lost from the system in the form of wind at a rate $\sim 10^{-4}\ M_{\sun}\ {\rm yr}^{-1}$, suggesting that the compact object in SS433 is accreting at a highly supercritical rate (see \citet{2004ASPRv..12....1F} for a detailed review on both observational and theoretical progress, and \citet{2019arXiv190502938C} for a more recent update). The binary orbital period, $P_{\rm orb}=13.1$ d, is well determined \citep{1981ApJ...251..604C}, but the nature of the compact object and the masses of the two components remain under debate. Spectroscopy studies have reported various mass estimates by constraining the mass function or the total mass of the binary  \citep[e.g.,][]{2008ApJ...676L..37H,2008ApJ...678L..47B,2010ApJ...709.1374K,2018MNRAS.479.4844C,2018A&A...619L...4B,2019MNRAS.485.2638C}. While the derived mass of the compact star spans a large range and some of them are in conflict with each other, most of the measurements reveal a mass larger than $3\,M_\sun$, hence pointing to a black hole (BH) rather than a neutron star (NS).

Another remarkable feature is that SS433 is in association with W50, a nebula resembling a supernova remnant (SNR). If it is indeed a SNR, as we will assume here \citep[for an alternative view, see][]{1980ApJ...238..722B}, it resulted from the formation of the BH. Hydrodynamical simulations on the formation of W50 and its interaction with SS433's jets have been employed to estimate the age of W50, and different studies consistently constrain it to be $\sim 10^4-10^5\ {\rm yr}$ \citep{S80,Z80,vdH80,L07,2011MNRAS.414.2838G,2017A&A...599A..77P}, and the duration of the current jet activity was estimated to be $\sim 10^3-10^4$ yr \citep{2000ApJ...530L..25K,2010arXiv1006.5213B,2011MNRAS.414.2838G}. Assuming W50 to be a SNR, SS433 is among the youngest XRBs, which include SXP1062 \citep{Henault2012}, Cir X-1 \citep{Heinz2013},  XMM JJ0513-6724 \citep{Maitra2019}, and SXP1323 \citep{Gvara2019}. These estimates also demonstrate that Roche-lobe overflow (RLOF) of the companion star commenced shortly after the BH was born, that is, the companion should be very close to filling its RL after the SN.

The supercritical mass transfer in SS433 is conventionally explained under the assumption that the donor is more massive than or of comparable mass to the BH and has a radiative envelope, so mass transfer proceeds on a thermal timescale \citep{2000ApJ...530L..25K}. Only a fraction of the transferred mass is deposited on the BH, with the rest expelled from the system, probably in the so-called ``re-emission" mode \citep{1999ApJ...519L.169K,2017MNRAS.471.4256V}.

The supercritical accretion distinguishes SS433 from other conventional XRBs, and its association with W50 may help constrain the properties of its progenitor system. Based on these characteristics, we use both binary population synthesis (BPS) and stellar evolution model to study the formation of SS433, in particular to explore why and how RLOF can start promptly after the SN. Our work is restricted to the case of BH binaries. In Section \ref{sec:bps} we describe our method and results on the formation of SS433 with the BPS calculation. A typical example of the mass transfer process is shown in Section 3 with detailed binary evolution calculation. We summarize and discuss their implications in Section \ref{sec:result}.

\section{Binary population synthesis}\label{sec:bps}
We first introduce our BPS model and then present the calculated results. The initial conditions of the primordial binaries and the input physics in the fiducial model are described in Section \ref{sec:bps_initial}. We also construct several alternative models to examine the influence caused by the uncertainties in massive star evolution and treatments in the accretion processes in Section \ref{sec:bps_models}. The results of all BPS models are elaborated and compared in Section \ref{sec:bps_result}.

\subsection{The fiducial model and initial parameters}\label{sec:bps_initial}
The picture for the formation of SS433 is quite simple and clear. We consider a primordial binary consisting of a more massive primary star (of mass $M_1$) and a less massive secondary star (of mass $M_2$, also called the companion star). The primary star evolves more rapidly, so it first expands and overflows its RL. The resultant mass transfer causes the binary orbit to shrink, and it could be dynamically stable or unstable depending on the evolutionary state of the primary star and the mass ratio. In the latter case a common envelope evolution (CEE) phase may ensue. After the mass transfer the primary star evolves to be a Helium star and finally explodes with a SN, giving birth to a BH. Under specific circumstances the secondary star can overflow its RL and transfer mass to the BH shortly after the SN. The goal of our work is to explore how the mass transfer proceeds and what are the requirements of the binary parameters.

We employ the BPS code originally developed by \citet{2000MNRAS.315..543H,2002MNRAS.329..897H} to follow the formation of the BH binary, where some treatments regarding the formation and evolution of massive binaries have been modified and updated \citep{2014ApJ...796...37S,2018MNRAS.477L.128S}. We briefly recapitulate the most relevant input physics here. (1) The wind mass loss scheme for luminous blue variables (LBVs) is the same as in \citet{2010ApJ...714.1217B}; for Wolf-Rayet stars, we follow the suggestion in \citet{2006MNRAS.369.1152K} and use half of the wind mass loss rate given in \citet{1998A&A...335.1003H}. (2) The critical ratios $q_{\rm cr}$ to distinguish stable and unstable mass transfer was evaluated in the original code with a semi-analytic way. Here we use the numerically calculated results presented in \citet{2014ApJ...796...37S} and \citet{2018MNRAS.477L.128S} in the cases of normal star accretor and BH accretor, respectively.

The initial binary parameters are set as follows. Following \citet{2002MNRAS.329..897H} we assume that there are 7.608 binaries with $M_1>0.8\ M_\sun$ born in the Galaxy per year. The distribution of the primary's mass obeys an initial mass function (IMF) $f(M_1) \propto M_1^{-2.7}$ \citep{1993MNRAS.262..545K}, and the mass ratio, $q=M_2/M_1$, has a flat distribution in $[0,1]$.
The logarithm of $M_1/M_{\sun}$, $M_2/M_{\sun}$, and the binary separation $a/R_{\sun}$  each has a flat distribution in $[\log(5),\ \log(100)]$, $[\log(1),\ \log(100)]$ and $[\log(3),\ \log(1000)]$, respectively.  We exclude those binaries whose initial total mass is smaller than $10\ M_{\sun}$ and evolve $1.5\times 10^7$ binaries for each constructed model. Evolution begins with both stars in zero-age main sequence (ZAMS) in a circular orbit, and terminates when the time $t$ reaches 10 Gyr or the binary is disrupted.

We adopt a CEE efficiency factor $\alpha=0.1$ in the BPS program. We also perform calculations with $\alpha=0.2,\ 0.5,\ 1.0$ following the conventional approach of parameter study, and find that variation in $\alpha$ does not cause a considerable change in either the number of the selected systems or their distribution. The reason is that most of the BH binaries we are concerned with are not expected to experience CEE prior to the second RLOF phase.

\subsection{Model considerations}\label{sec:bps_models}
There are some uncertainties in the BPS studies that can significantly influence the formation of SS433-like systems. The most relevant ones are as follows.

1. The mass transfer efficiency during the binary star evolution. When mass is transferred from the primary star to the secondary star, how much is accreted and how much is lost from the binary? These processes are ill-constrained but influence the evolution prior to the birth of the BH. To include different possibilities, we adopt the approach in \citet{2014ApJ...796...37S} considering three modes with increasing mass transfer efficiency, from mode I with rotation-dependent, highly non-conservative accretion,  mode II with half mass accretion and half mass loss, to mode III with thermal equilibrium-limited, nearly conservative accretion.

2. The remnant mass and the natal kick distribution produced by the SN. Current investigations on the mechanisms of SN explosions still cannot make precise predictions on the nature and the mass of the compact object from the progenitor mass, so we try two prescriptions when determining the remnant mass. One is based on the rapid SN mechanism suggested in \citet{2012ApJ...749...91F}, which may account for the observed $\sim 3-5\,M_\sun$ gap in the NS/BH mass distribution. The other assumes that the pre-explosion helium core may entirely go into the BH for stars
with initial masses of $15-40\,M_\sun$ \citep[e.g.,][]{Sukhbold2016}\footnote{The average likelihood for this kind of implosion was estimated to be $\sim 57.4\%$ to fit the observed NS/BH mass distribution \citep{R18}.}. The amplitude of the natal kick is assumed the follow the distribution the NS kick velocities \citep{Hobbs2005}, but reduced by multiplication by a factor of $(1 - f_{\rm fb})$ where $f_{\rm fb}$ is the SN fallback fraction, or $3M_\sun/M_{\rm BH}$ where $M_{\rm BH}$ is the BH mass, in these two prescriptions, respectively.

The details of our fiducial model A1 and its variations are summarized in Table \ref{tabel:1}.

\begin{table}[h]
\centering
\caption{Description of the binary population synthesis Models.}
\begin{tabular}{lllll}
\hline
Model   & BH mass prescription and natal kick                                   & mass transfer efficiency       &  &  \\
\hline
A1 & \citet{2012ApJ...749...91F}, NS kick modified according to $f_{\rm fb}$   & rotation-dependent             &  &  \\
A2 & same as in A1                                                 & 0.5                 &  &  \\
A3 & same as in A1                                                 & limited by thermal equilibrium &  &  \\
B1 & mass of pre-SN Helium core, NS kick normalized by BH mass & same as in A1                          &  & \\
\hline
\end{tabular}

\label{tabel:1}
\end{table}

\subsection{Selection criteria and calculated results}\label{sec:bps_result}
\subsubsection{The initial BH binaries in the fiducial model}\label{sec:bps_select}
To obtain BH binaries that are associated with an observable SNR, we search the BH binaries in our BPS results whose XRB phase starts within $10^5$ yr since the birth of the BH, i.e., $t_{\rm BH-XRB}<10^5$ yr. To aid further analysis, we define a RL-filling factor of the companion star, $f=R_2/R_{\rm L2}$, where $R_2$ and $R_{\rm L2}$ are the radius and the RL radius of the companion star, respectively. We set a binary to enter or exit the XRB phase when $f$ becomes larger or smaller than $0.95$, respectively. The BH binaries with $f$ initially larger than 2.0 are excluded as this implies that the BH is already embedded in the companion's envelope. We let $t_{\rm BH}$ be the birth time of the BH, and $t_1$ and $t_2$ the beginning and the end of the RLOF phase, respectively. With these values defined, the time spent from the BH's birth to the onset of RLOF can be expressed as  $t_{\rm BH-XRB}=t_1-t_{\rm BH}$ and the duration of the RLOF phase $t_{\rm XRB}=t_2-t_1$.

According to the observations of SS433 we require that the average mass loss rate $\dot{M}_2$ of the donor to be larger than $\sim 10^{-8}\ M_{\sun}\ {\rm yr}^{-1}$.
We restrict the companion mass to be less than $50\,M_\sun$ at the birth of the BH and the orbital period $P_{\rm orb}$ at the beginning of RLOF shorter than one year. The initial BH binaries with $P_{\rm orb}> 1$ yr are excluded for two reasons: (1) The extent of orbital shrinkage caused by RLOF is limited and hence a realistic binary is unlikely to evolve from $P_{\rm orb}>1$ yr to the currently observed $P_{\rm orb}=13$ d of SS433 without extreme mass ratio ($M_{\rm BH}/M_2<0.2$); (2) the RL size of the companion star in a 1 yr orbit is approximately 300 $R_\sun$ (with $M_{\rm BH}\sim 10\ M_\sun$ and $M_{\rm 2}\sim 20\ M_\sun$),  which means that the companion star should be significantly evolved, likely leading to dynamically unstable mass transfer.

To summarize, the initial BH binaries we select should satisfy the following criteria:\begin{enumerate}
\item $t_{\rm BH-XRB}<10^5$ yr;
\item $\overline{\dot{M}_{\rm 2}} > 10^{-8}\ M_{\sun}\ {\rm yr}^{-1}$;
\item $P_{\rm orb}$ at the beginning of RLOF $<1$ yr;
\item $M_{\rm 2}<50\ M_\sun$.
\end{enumerate}

\begin{figure}[!h]
	\centering
	\includegraphics[width=1.0\textwidth]{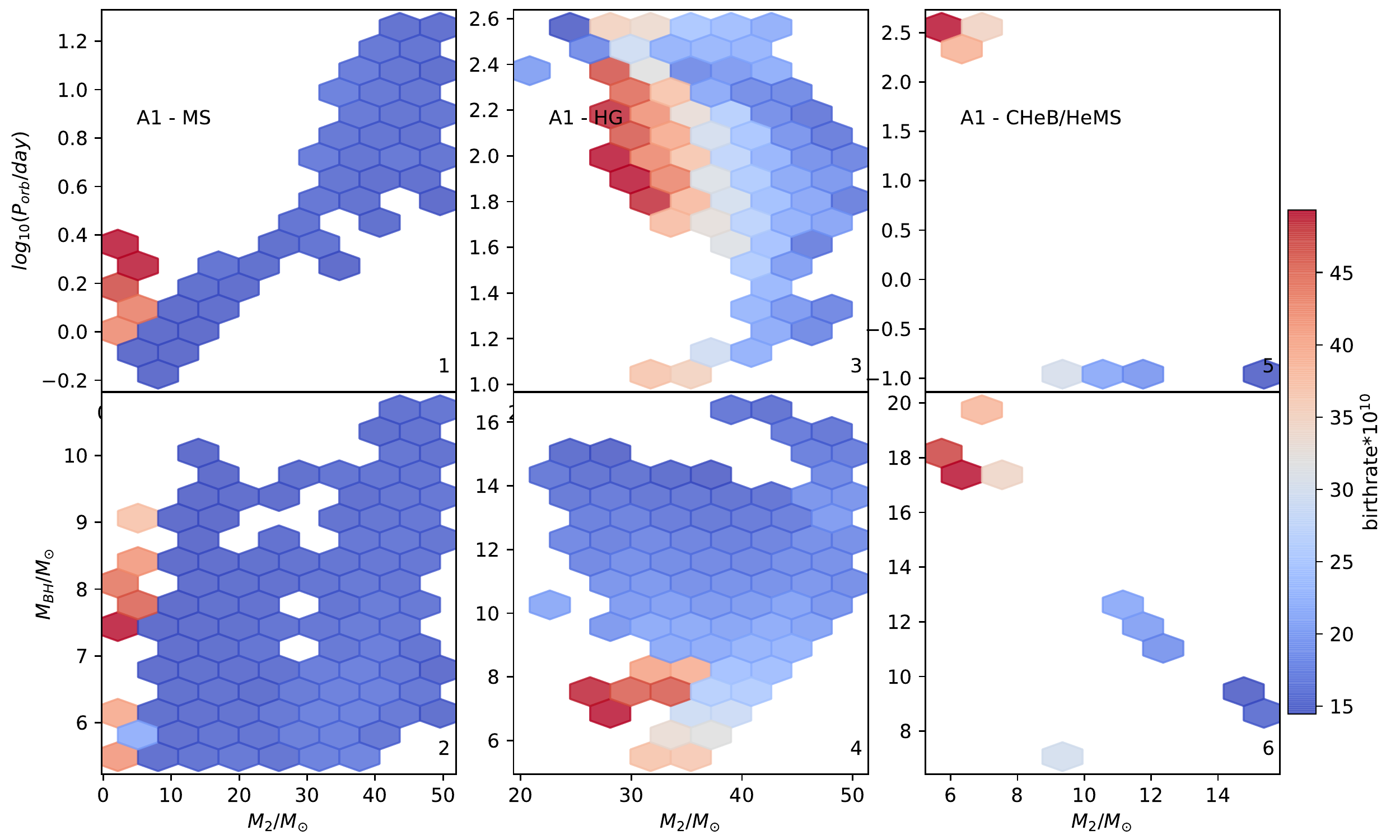}
	\caption{Distribution of the BH mass and the orbital period versus the companion mass for the initial BH binaries in model A1. The three columns correspond to three groups of BH binaries separated by the evolutionary stage of the companion star: (1) BH binaries with a MS donor, (2) BH binaries with a HG donor, and (3) BH binaries with a CHeB or HeMS donor.  The colors of the hexagons show the birthrate of the BH binaries in the corresponding parameter space amplified by a factor of $10^{10}$.
}\label{fig:Fig1}
\end{figure}

Figure 1 demonstrates the distributions of the companion mass, BH mass, and orbital period at the beginning of RLOF for the fiducial model A1. In the following we will compare them with those for other models. Since the criteria listed above are rather loose, the properties of the selected binaries span a wide range and we divide them into three groups according to the evolutionary stage of the companion star: (1) BH binaries with a MS companion, (2) BH binaries with a Hertzsprung gap (HG) companion, and (3) BH binaries with a core Helium burning (CHeB) star or Helium main sequence (HeMS) star companion. They are displayed in Fig.~\ref{fig:Fig1} with hexagons, where the BH/MS binaries are in the left panels (panels 1 and 2), BH/HG binaries in the middle panels (panels 3 and 4), and BH/CHeB/HeMS binaries in the right panels (panels 5 and 6). The upper and lower panels demonstrate their distributions in the $P_{\rm orb}-M_2$ and $M_{\rm BH}-M_2$ planes, respectively. We use different colors to denote the number of the birthrate for each binary, which is amplified by a factor of $10^{10}$. The overall birthrate of the BH binaries in model A1 is $R_{\rm BH-SNR, A1}=7.35\times 10^{-6}\ {\rm yr^{-1}}$.

We then separately discuss the properties and possible evolution of the three groups of BH binaries.

For all BH/MS binaries, $P_{\rm orb}<20$ d and $M_2$ varies from less than $10\ M_\sun$ to over $40\ M_\sun$, clustered in the region with $P_{\rm orb}<3$ d and $M_2<8M_\sun$. The BH mass ranges from $5M_\sun$ to $11M_\sun$. There seems to be roughly positive relationship between $P_{\rm orb}$ and $M_2$ in panel 1 of Fig.~\ref{fig:Fig1}, which actually reflects the mass-radius relation of MS stars. For example, a binary with a 13 d orbit  harbouring a 10 $M_\sun$ BH corresponds to a MS donor more massive than 40 $M_\sun$. The BH/MS binaries with mass ratio below the critical value will enter CEE \citep{2018MNRAS.477L.128S}, and others can proceed mass transfer steadily. The average mass transfer rate during the RLOF phase is dependent on the companion mass, varying between  $\sim 10^{-8}\ M_{\sun}\ {\rm yr}^{-1}$ and $\sim 10^{-5}\ M_{\sun}\ {\rm yr}^{-1}$.

The size of stars that have evolved off MS can change drastically, therefore the distribution of the BH binaries with evolved donors shows little correlation between $M_2$ and $P_{\rm orb}$, as depicted in panels 3 and 5 of Fig.~\ref{fig:Fig1}. The HG donors usually have expanded their radii to $\gtrsim 100 R_\sun$ when RLOF onsets. Most of the HG donors are still on the MS at the birth of the BH, and the subsequent RLOF is facilitated by the rapid evolution of these MS stars - they are massive enough to reach the late stage of MS when the BH is born and spend less than $10^5$ yrs evolving off MS after that. Specifically, for the companion to fill its RL in a $P_{\rm orb}>20$ day orbit within $10^5$ yrs since the BH's birth, the companion mass has to be $\sim 30\,M_\sun$. The mass ratios of the BH/HG binaries are systematically higher than that of BH/MS binaries, so a larger fraction of them are able to avoid CEE, and the mass transfer rate is usually as high as $\sim 10^{-3}\ M_{\sun}\ {\rm yr}^{-1}$.

The BH binaries with even more evolved donors are plotted in the right panels. They have either very short($<0.1$ d) or long ($>100$ d) orbital periods, corresponding to HeMS and CHeB donors, respectively. On the one hand, the BHs with CHeB donors are of relatively  low-mass ($M_{\rm BH} \leq 10 M_\sun$). This is because lower-mass BHs evolved from less massive primaries with a longer lifetime, which allows the secondary to ``catch up" and evolve to the CHeB phase at the BH's birth. The RLOF phase generally lasts $\lesssim 1000$ yrs with the mass transfer rate $\dot{M}_{2}\sim 10^{-5}\ M_{\sun}\ {\rm yr}^{-1}$. On the other hand, the HeMS donors are usually paired with relatively massive BHs ($M_{\rm BH} \geq 10 M_\sun$) in compact orbits. They most probably have experienced CEE prior to the RLOF phase. The mass transfer phase lasts for over $10^5$ yrs at a mean rate  $\sim 10^{-6}\ M_{\sun}\ {\rm yr}^{-1}$.

In summary, compared with the properties of SS433, we find that the BH/MS binaries have a too massive donor star, the BH/CHeB binaries have a too short mass transfer phase, and the BH/HeMS binaries have a too small orbit. This leaves the BH/HG binaries to be the most promising candidate as the progenitor of SS433.

\begin{figure}[!h]
	\centering
	\includegraphics[width=1.0\textwidth]{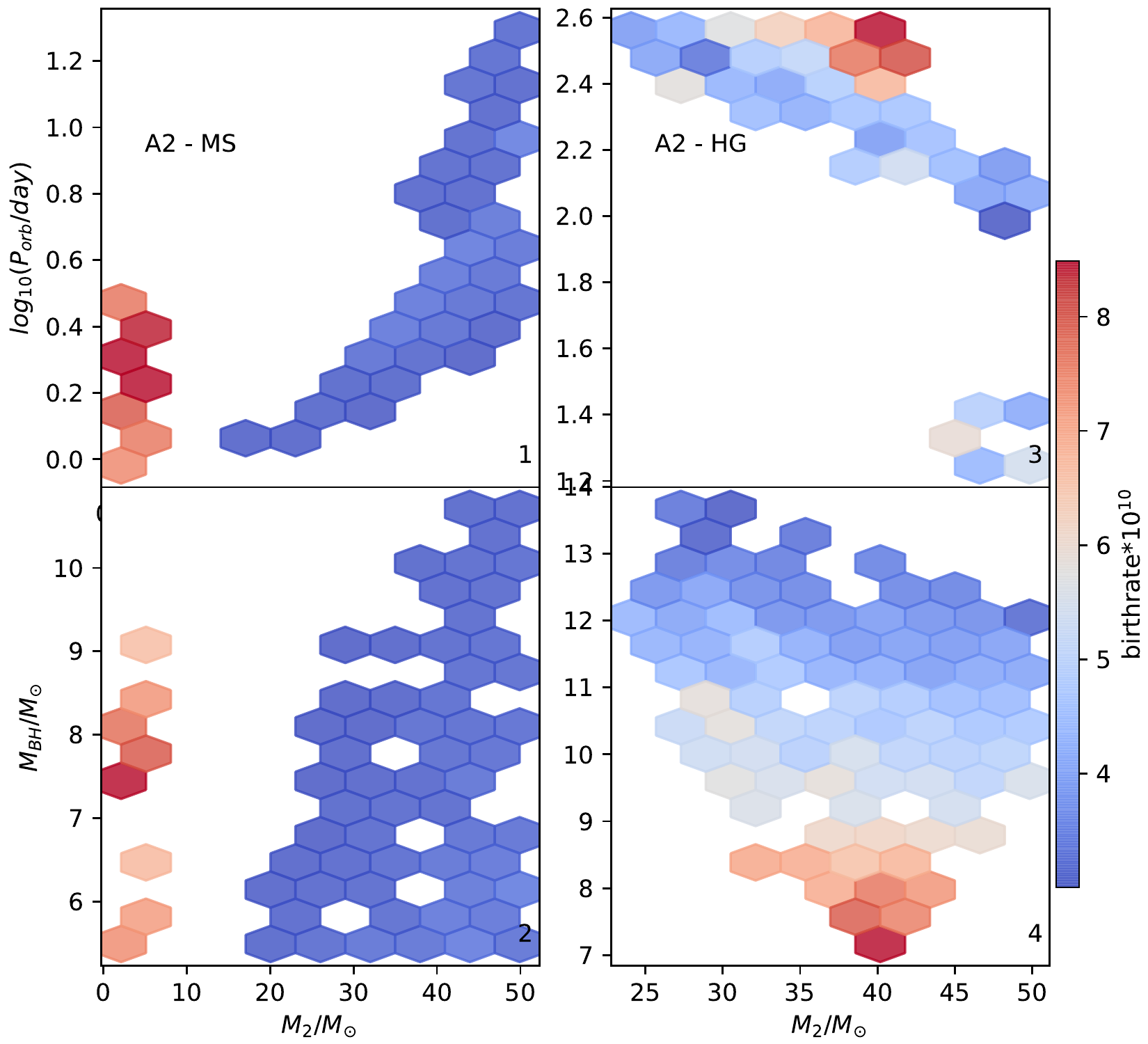}
	\caption{Same as Fig.~1 but for model A2. The left and right columns correspond to BH/MS and BH/HG binaries, respectively.
	}\label{fig:Fig2}
\end{figure}

\begin{figure}[!h]
	\centering
	\includegraphics[width=1.0\textwidth]{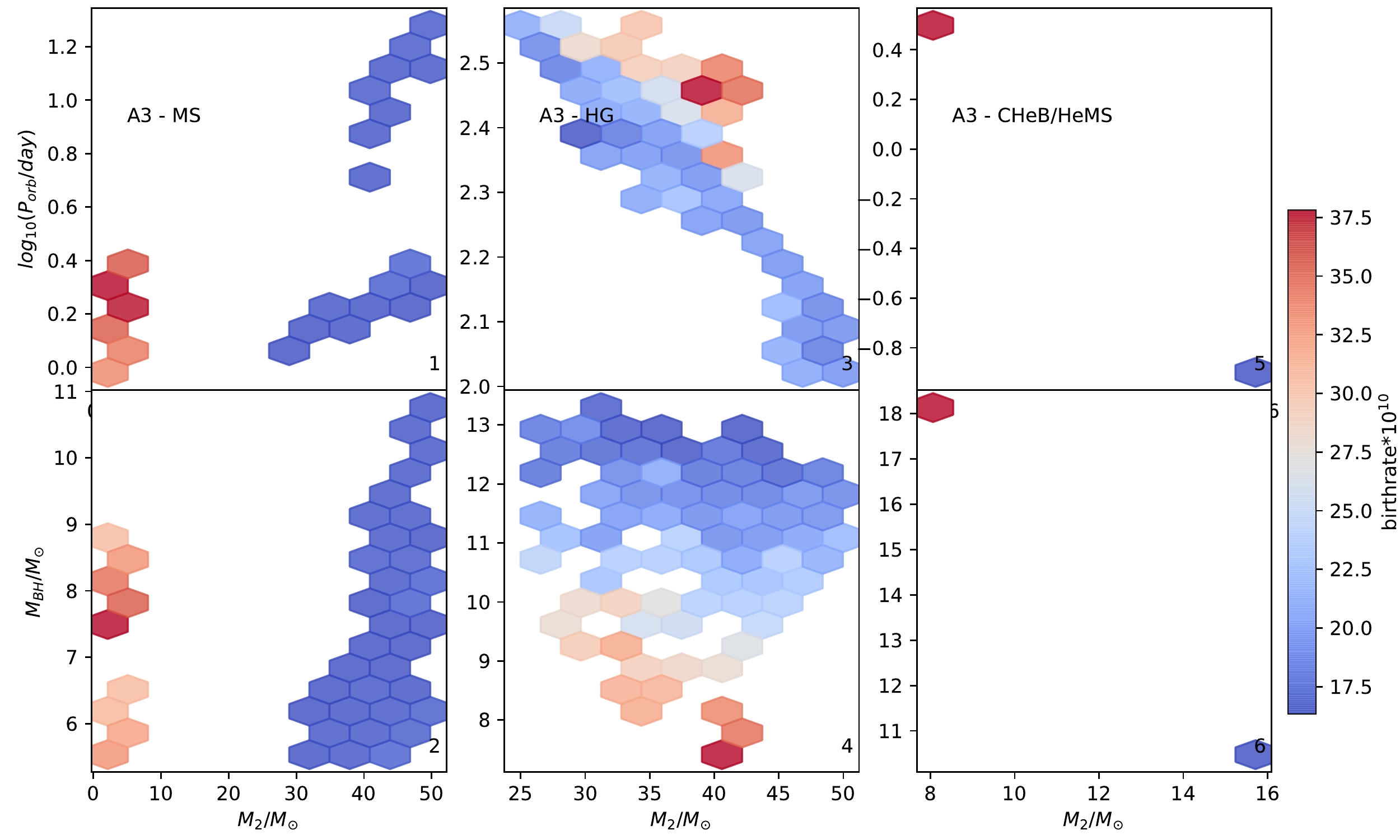}
	\caption{Same as Fig.~1 but for model A3.
}\label{fig:Fig3}
\end{figure}

\begin{figure}[!h]
	\centering
	\includegraphics[width=1.0\textwidth]{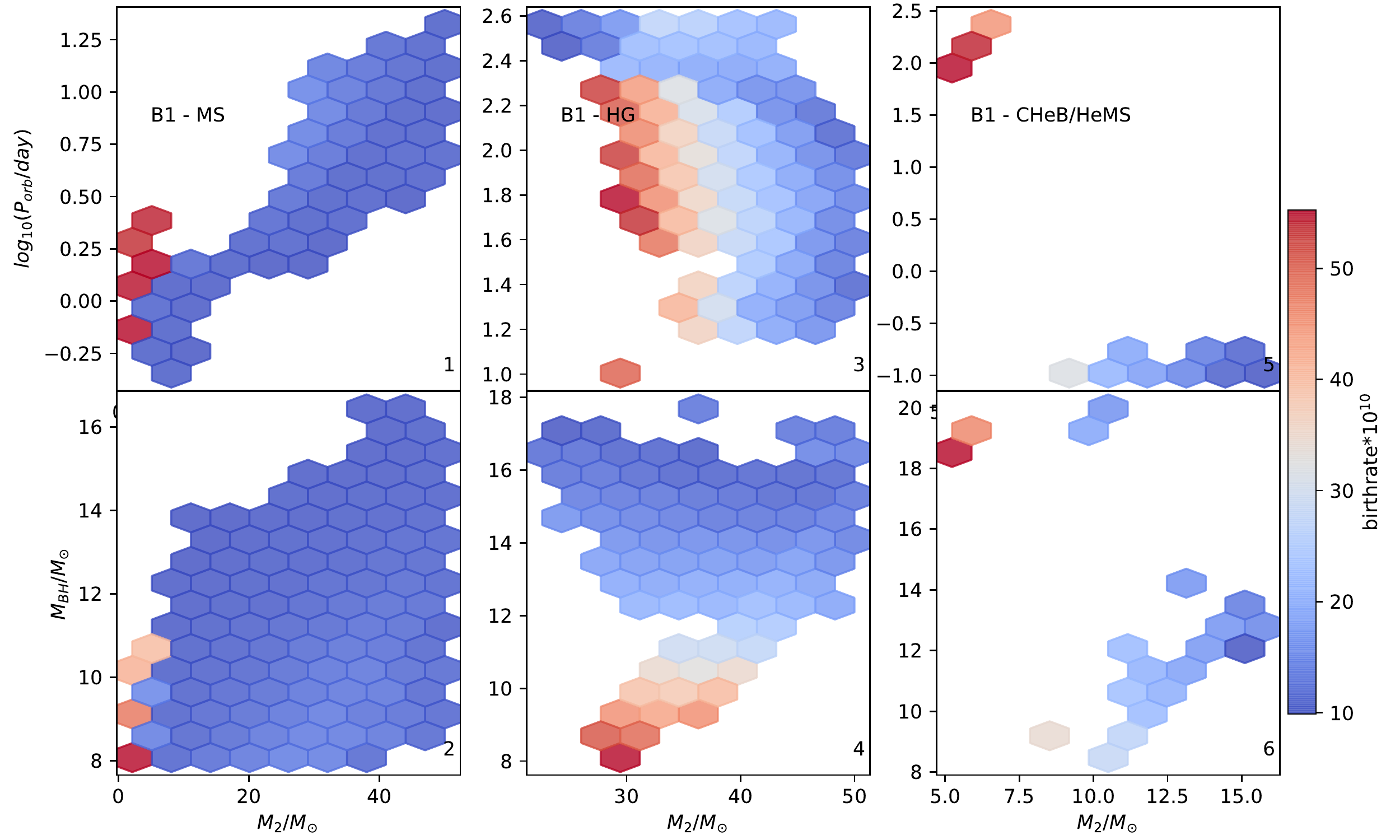}
	\caption{Same as Fig.~1 but for model B1.
	}\label{fig:Fig4}
\end{figure}

\subsubsection{Comparison with other models}

We then investigate the possible influence of different mass transfer modes and BH mass prescriptions.

The results of models A2 and A3 are shown  in Figs.~\ref{fig:Fig2} and \ref{fig:Fig3}, respectively. The selection criteria are the same as those for model A1 as listed in Section~\ref{sec:bps_select}. Their differences lie in that the fraction of the accreted mass during the first mass transfer process increases from model A1 (highly nonconservative) to model A3 (nearly conservative), so the secondary star is rejuvenated by a different amount in these models. Because the secondary has accreted more mass from the primary in models A2 and A3 than in model A1, the evolution of the primordial binaries with exactly the same initial configuration can be quite different. In models A2 and A3 they will evolve into BH binaries with a more massive companion star and more likely violate our selection criteria.

There is a deficiency in BH/MS binaries with companion mass $M_2\sim 10\ M_\sun-30\ M_\sun$ in models A2 and A3 compared with the fiducial model, as shown in panel 1 of Figs.~\ref{fig:Fig2}  and \ref{fig:Fig3}.
By comparing panel 3 of Figs.~\ref{fig:Fig1}-\ref{fig:Fig3}, we can see that the BH/HG binaries with moderate-mass ($M_2\sim 20-30\,M_\sun$) companions and relatively long orbital period ($P_{\rm orb}\sim100-200$ d) are absent in models A2 and A3. This parameter space corresponds to the BH/HG binaries that have experienced substantial orbital shrinkage in models A1, which are the most promising to evolve into systems with similar observational properties of SS433. As for the BH/CHeB/HeMS binaries, we found none in model A2 and much fewer of them in model A3 than in model A1. These cause a decrease in the overall birthrate of BH binaries in models A2 and A3 ($R_{\rm BH-SNR, A2}=2.18\times10^{-6}\ {\rm yr^{-1}}$ and $R_{\rm BH-SNR, A3}=2.07\times10^{-6}\ {\rm yr^{-1}}$).

Then we compare the results of models A1 and B1 to explore the influence of the BH mass prescription. The BH mass in model B1 equals the mass of the pre-explosion He core, larger than that in model A1, which is essentially determined by the pre-explosion CO core mass. The binary evolutions prior to the birth of the BH in the two models are almost identical, therefore their distributions on the $M_{\rm BH}-M_2$ plane are very similar. The three groups of binaries identified in model A1 are all present in model B1 but with slightly different distributions. There is a systematic shift in the $M_2-M_{\rm BH}$ distribution due to the increase in the mass ratio for all three groups of the BH binaries, as shown in lower panels of Figs.~\ref{fig:Fig1}  and \ref{fig:Fig4}. The kick velocity of the BHs in models A1 and B1 is scaled according to the fallback mass fraction and the BH mass, respectively. Most of the selected binaries received a small kick (less than $50\,{\rm km\ s^{-1}}$), hence the difference in the orbital properties caused by the kick velocity is insignificant, and there is little difference in the orbital period distribution in the upper panels of Figs.~\ref{fig:Fig1} and \ref{fig:Fig4}.
The birthrate of model B1 is, as expected, close to that of the fiducial model, $R_{\rm BH-SNR, B1}=6.60\times10^{-6}\ {\rm yr^{-1}}$.

\subsubsection{Further constraints for SS433}\label{sec:bps_ss}

For the individual system SS433, more specific constraints are added according to its well determined orbital period and estimated age. Specifically, we select BH binaries with $t_{\rm BH-XRB}<10^4$ yr, i.e., the time the companion spent to fill its RL after the BH's birth is less than $10^4$ yrs. We also require that the orbital period is shorter than 20 days at the end of the RLOF phase. Additionally, we put a constraint on the duration of the RLOF phase $t_{\rm XRB}\geq 10^3$ yr. BH/MS binaries are unlikely to be the progenitor systems of SS433 because the donor mass is too large, in conflict with the observed properties of the optical star. We find that all the binaries that satisfy these more stringent criteria have HG donors at the beginning of the RLOF. They can be considered as a refined sample of the BH/HG binaries selected as the BH-SNR candidates in the last subsection. As we pointed out before, there is a deficiency of this kind of binaries in models A2 and A3. Consequently, these two models are unable to reproduce the BH binaries that satisfy the criteria for SS433.

\begin{figure}[!h]
	\centering
	\includegraphics[width=1.0\textwidth]{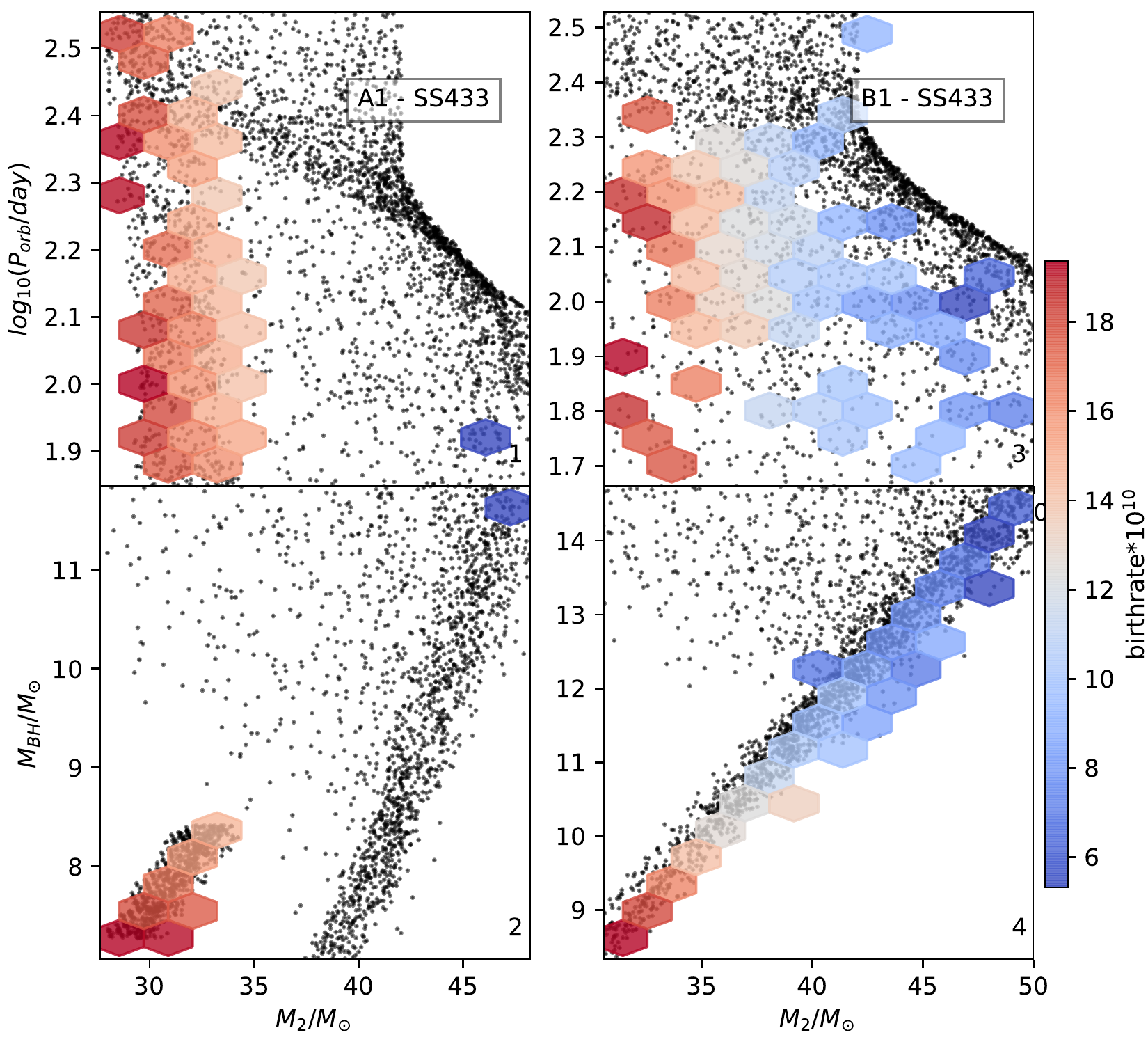}
	\caption{The hexagons show the distribution of the BH mass and the orbital period versus the companion mass for the BH binaries that will evolve to be SS433. The left and right panel correspond to models A1 and B1, respectively. The black dots that scatter over the background represent the BH/HG binaries in Figs.~1 and 4 within the parameter space of this figure.
	}\label{fig:Fig5}
\end{figure}

Figure~\ref{fig:Fig5} shows the distributions of the BH mass, companion mass, and orbital period in models A1 (left panels) and B1 (right panels) with hexagons for the thus constrained BH/HG binaries. We also plot all the BH/HG binaries displayed in Figs.~\ref{fig:Fig1} and \ref{fig:Fig4} with black dots as the background. From the contrast between the hexagons and the black dots one can see which systems are removed when more strict selection criteria are applied. For example, for the fiducial model A1, binaries whose mass ratio is too small are excluded because they will soon enter CEE, and those whose mass ratio is too large are also excluded because they do not experience substantial orbital shrinkage during the RLOF phase. This results in the fact that the suitable binaries are distributed in a rather small parameter space in the  $M_{\rm BH}-M_2$ plane.
For model B1, the change in the mass distribution is less discernible due to the increased BH masses. The binaries with relatively large mass ratio or long orbital period are removed for the same reason as explained for model A1. Binaries with massive donors are more abundant in model B1 as the BH masses are larger. The birthrates of the BH XRBs in models A1 and B1 are $R_{\rm SS433, A1}=1.32\times10^{-7}\ {\rm yr^{-1}}$ and $R_{\rm SS433, B1}=1.04\times10^{-7}\ {\rm yr^{-1}}$, respectively.

\begin{figure}[!h]
	\centering
	\includegraphics[width=1.0\textwidth]{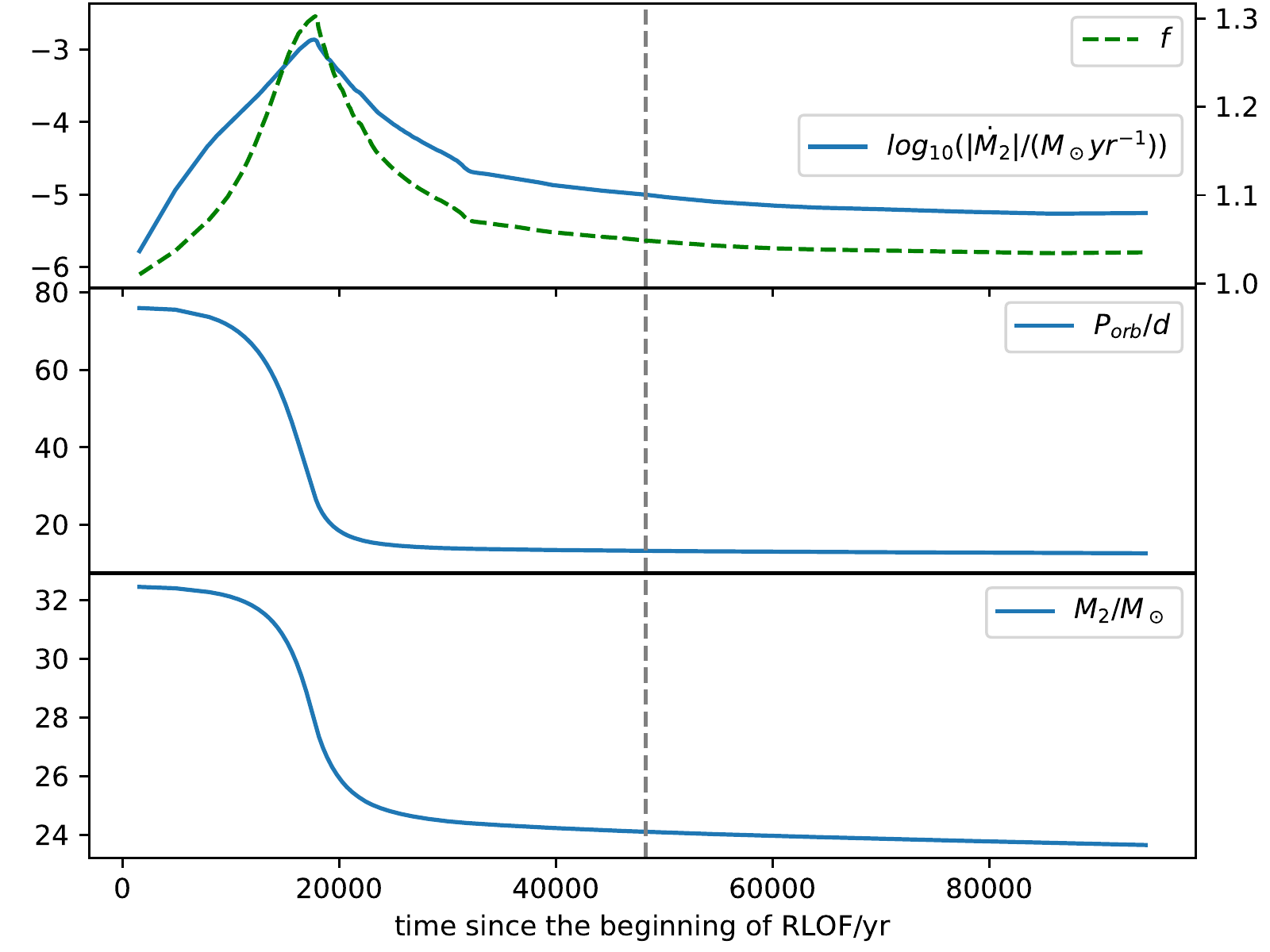}
	\caption{An example of the binary evolution that may lead to the formation of SS433-like systems. Shown are the mass transfer rate (with the solid line in the upper panel), the RL-filling factor of the companion (with the dashed line in the upper panel), the orbital period (middle panel), and the companion mass (lower panel) as a function of time. The vertical dashed line denotes the time ($4.67\times 10^4$ yr) when $P_{\rm orb}=13$ d. At this time the masses of the BH and the companion star are $M_{\rm BH}=$ 7.92 $M_\sun$ $M_2=$24.12 $M_\sun$, respectively.
	}\label{fig:Fig6}
\end{figure}



\section{An example of the mass transfer process}\label{sec:mesa}
The detailed mass transfer process that starts after the BH's birth is poorly treated with BPS, because the code cannot evaluate the variation of the donor radius in response to mass loss, and the calculated mass transfer rates based on the RL-filling factor $f$ are not sufficiently accurate. We therefore use the stellar evolution and the binary modules in Modules for Experiments in Stellar Astrophysics (MESA) \citep{2011ApJS..192....3P,2013ApJS..208....4P,2015ApJS..220...15P,2018ApJS..234...34P,2019ApJS..243...10P} to follow the binary evolution after the mass transfer.

We first construct a model to describe the binaries at the beginning of RLOF, where only three parameters, $M_{\rm BH}$, $M_{2}$, and $P_{\rm orb}$ are specified. We let the mass of a ZAMS star increase by a certain amount of mass, so that when it evolves off the MS and expands to fill its RL, its mass is close to $M_{2}$. This approximation is plausible considering that the secondary star was only slightly rejuvenated in models A1 and B1. Therefore, the stellar structure of the HG stars in the selected sample from the BPS study does not considerably deviate from that of the HG stars we evolve with MESA. Then this HG star is put in a binary orbit along with a point mass which represents the BH. We use the \citet{1990A&A...236..385K} scheme in MESA to calculate the  mass transfer rate under optically thick circumstances during the RLOF phase. The time is set to be zero at the beginning of the mass transfer so that the age of the binary in MESA corresponds to the time it spent since the onset of RLOF.

A typical example of the binary evolution is shown in in Fig.~6. The initial parameters of this binary are specified with $M_2=$ 32.45 $M_\sun$,  $M_{\rm BH}=$7.90 $M_\sun$, and $P_{\rm orb}=$75.99 d. We plot the time evolution of the donor mass, orbital period, mass transfer rate and RL-filling factor. The top panel shows that the mass transfer rate is closely related to the RL-filling factor $f$, and for this particular binary, $f$ can reach 1.3 but stays below 2.0 through the whole mass transfer phase. There is an episode of approximately several thousand years when the mass transfer rate can be as high as $10^{-3}\ M_{\sun}\,{\rm yr}^{-1}$, but the average mass transfer rate is approximately $10^{-5}\,M_{\sun}\,{\rm yr}^{-1}$, significantly lower than that ($\sim 10^{-3}\,M_{\sun}\,{\rm yr}^{-1}$) in the BSE calculation. The orbital period decreases rapidly during the first $2\times 10^4$ yr along with mass transfer. The grey vertical dotted line marks the time when the orbital period has reduced to 13.1 d. At this time the BH mass and the companion mass are $M_{\rm BH}=$ 7.92 $M_\sun$ and $M_2=$24.12 $M_\sun$, respectively. If we consider that $P_{\rm orb}<20$ d is an acceptable requirement for the orbital period, then the evolution can well reproduce the measured mass transfer rate and age of SS433.

A problem with this evolution is that it is difficult to account for the spectral type of the companion star. We plot the evolutionary track of the companion star in the H$-$R diagram (Fig.~7). The red cross marks the time of $P_{\rm orb}=13.1$ d, and the corresponding effective temperature is $T_{\rm eff}=2.4\times 10^4$ K. This value is higher than the observed value  in previous studies, in which the spectral type of the optical star was suggested to be A4I-A8I \citep{2004ApJ...615..422H,2008ApJ...676L..37H,2010ApJ...709.1374K,G11}. These observations indicate that the mass of the optical companion star falls within the range $\sim 8.3-12.5\,M_\sun$, with an upper limit of $15\,M_\sun$, considerably lower than our estimate. However, we note that the mass of the companion star is highly controversial in the literature due to the difficulty in identification of the absorption lines. There are also reports on the companion mass $\gtrsim 25-40\,M_\sun$ \citep{L06,C13,2019MNRAS.485.2638C}.

\begin{figure}[!h]
	\centering
	\includegraphics[width=0.6\textwidth]{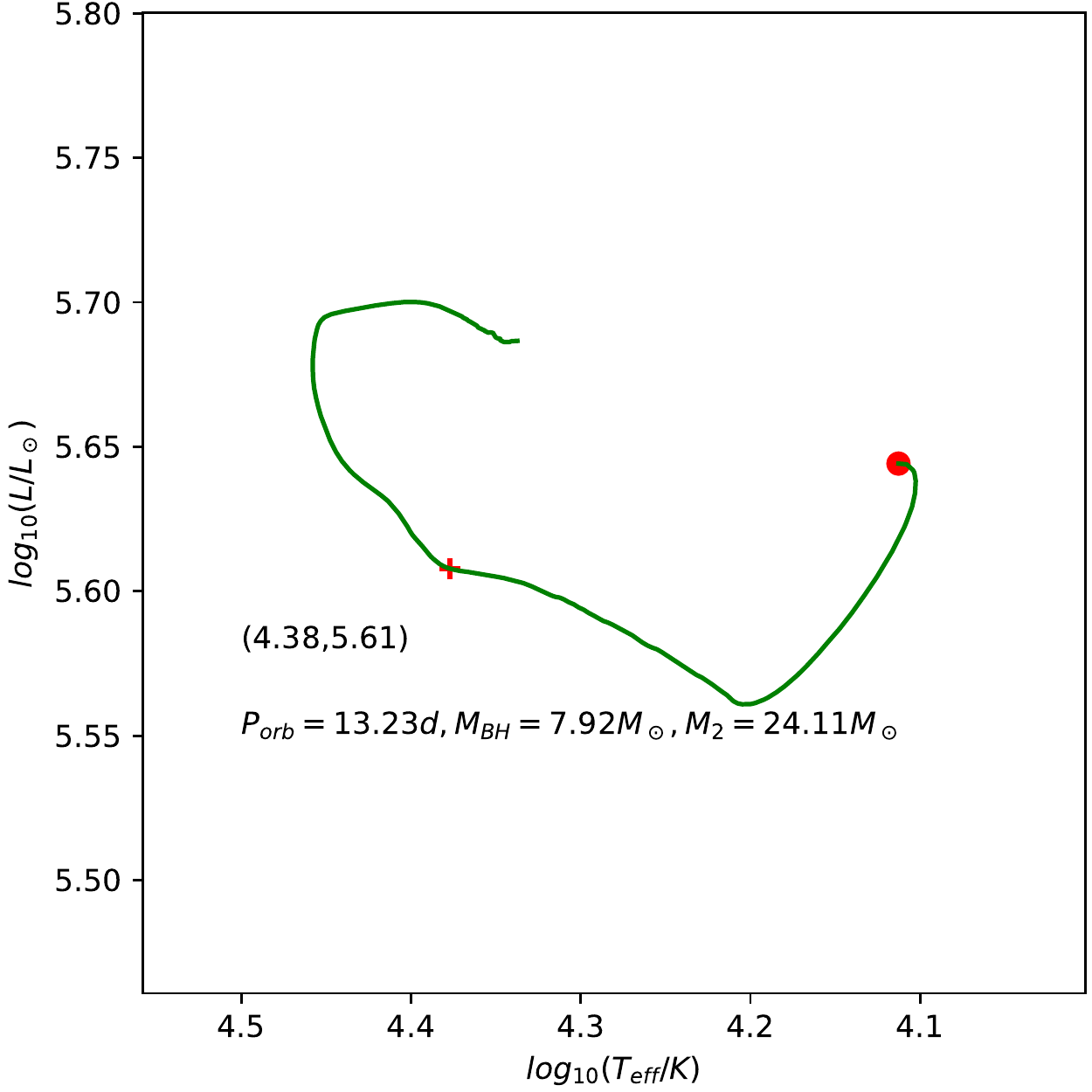}
	\caption{The evolutionary track of the companion star in H$-$R diagram during the mass transfer phase. The red dot marks the beginning of RLOF. The red cross denotes the time when $P_{\rm orb}=13$ d. The values in parenthesis denote the logarithm of the effective temperature (in units of K) and the the luminosity (in units of $L_\sun$) of the companion star.}\label{fig:Fig7}
\end{figure}



\section{Discussion and Conclusions}\label{sec:result}

By BPS calculations we have explored the possible progenitors of BH XRBs with very young age. Combining the results presented in Figs.~1-4, we find that BH binaries whose RLOF phase commenced shortly after the BH's birth can possess MS, HG, CHeB and HeMS companion stars.  Their orbital periods at the onset of RLOF range from less than 1 d to several hundred days. The birthrate  in the Galaxy is approximately $10^{-6}\ {\rm yr^{-1}}$ in our fiducial model. However, only a small fraction of these initial BH binaries have the opportunity to evolve into SS433-like systems, because the mass transfer processes sensitively depend on the mass ratio and the evolutionary state of the companion stars. BH/HG binaries most likely satisfy the requirements (age, donor mass and orbital period) for the formation of SS433. The mass and orbital period at the onset of RLOF are $M_2\gtrsim 25-30\ M_\sun$ and $P_{\rm orb}\gtrsim 50-80$ d. A typical example of binary evolution suggests the the current masses of the BH and the optical star to be $M_{\rm BH}\sim 8\ M_\sun$, $M_{\rm 2}\sim 24\ M_\sun$, respectively. The corresponding birthrate is around $10^{-7}\ {\rm yr^{-1}}$ in models A1 and B1. Considering the fact that the birthrate of BH binaries in the Galaxy is a few $10^{-5}-10^{-4}\,{\rm yr^{-1}}$ \citep{SL19} and there is only one such X-ray source (i.e., Cygnus X-1) discovered in the Galaxy, the very low birthrate of SS433 suggests that either we are extremely lucky to observe such a rare object or the actual age of SS433 is much longer than $10^5$ yr (see below).

There are several sources of uncertainties in our BPS method.
Regarding the evolution of massive stars, the uncertainties in the wind loss scheme and the prescription for determining the mass of the compact object inevitably cause quantitative difference in the masses of the two components and the orbital period at the beginning of the RLOF phase. Nevertheless, we argue that our BPS results strongly indicate a massive secondary ($M_2\gtrsim 30\ M_\sun$) at the onset of the RLOF phase. The reason is rather straightforward: a secondary star with significantly lower mass has a much longer MS lifetime than the primary star, and hence is unable to expand and overflow its RL shortly after the BH's birth.

Furthermore, our selection criterion on $t_{\rm BH-XRB}$ (the time the companion star spent to fill its RL after the SN) is based on the estimation from hydrodynamic simulations of the SNR. Relaxing $t_{\rm BH-XRB}$ from $10^4$ yr to $10^5$ yr would decrease the lower limit of the donor mass. However, considering the evolution history of BH/HG binaries, this lower limit is also constrained by the requirement that mass transfer should help shrink an initially wide orbit ($P_{\rm orb}> 50$ days) to the current size ($P_{\rm orb}\sim 13$ days).

All of the BH/HG binaries in our simulation have a period of super-Eddington mass transfer, implying strong mass loss. Some of the mass could be lost through the L2/L3 points which is not taken into account. However, we do not expect it to significantly alter our results because the mass loss rate through the L2/L3 points is far smaller than that through the L1 point and only lasts for short duration \citep{2017MNRAS.465.2092P}.

Finally, we note that although the radio nebula W50 is generally considered as a SNR, its origin is still uncertain and debated \citep[][for a review]{F17}. Its elongated morphology is very likely caused by the interaction between SS433 and W50. Besides the SNR+Jets model which favors the interaction between the jets from SS433 and an already SNR \citep{Z80}, there is an alternative Wind+Jets model, which suggests the expansion of the SS433 jets or wind swept a bubble in the interstellar medium \citep{1980ApJ...238..722B}.
In that case the tight constraint on the time between the SN and the start of RLOF no longer exists. For example, if we change the upper limit of $t_{\rm BH-XRB}$ from $10^5$ yr to $10^6$ or $10^7$ yr, then the birthrate of the BH XRBs increases to $2.11\times 10^{-5}$ or $4.0\times 10^{-5}$ yr$^{-1}$, respectively. In this regard the identification of the optical companion star is crucial, because it can not only set stringent constraint of the formation history of SS433, but also help determine the nature of W50. If the companion star is a supergiant star with mass significantly less than $20\,M_\sun$, then it is unlikely that W50 was formed due to the SN explosion that gave birth to the BH.

However, there is a more exotic possibility that W50 is indeed a SNR but the accretion disk surrounding the BH and the circumbinary disk originated from the fallback material from the SN \citep{C71,C89} rather than from the companion star. In this scenario, the companion star is actually detached from its RL, and the stringent requirement of its rapid evolution also does not exist. Again this can be tested by future measurement of the mass and size of the companion star.

\acknowledgements
We are grateful to the referee for helpful comments. This work was supported by the National Key Research and Development Program of China (2016YFA0400803), the Natural Science Foundation of China under grant No. 11773015 and Project U1838201 supported by NSFC and CAS.

\end{document}